\documentclass[graybox]{svmult} 
\usepackage{mathptmx}       
\usepackage{helvet}         
\usepackage{courier}        
\usepackage{type1cm}        
                            
\usepackage{makeidx}         
\usepackage{graphicx}        
\usepackage{multicol}        
\usepackage[bottom]{footmisc}

\makeindex             

\usepackage{graphicx}
\usepackage[usenames,dvipsnames,svgnames,table]{xcolor}
\usepackage{amsmath}
\usepackage{amssymb}
\usepackage{setspace}
\usepackage{hyperref}
\usepackage{url}
\definecolor{light-gray}{gray}{0.8}

\usepackage{subfigure}

\begin{document}

\title*{Optimal modularity in complex contagion}%
\author{Azadeh Nematzadeh, Nathaniel Rodriguez, Alessandro Flammini, and Yong-Yeol Ahn}
\institute{Azadeh Nematzadeh \at Center for Complex Networks and Systems Research, School of Informatics and Computing, Indiana University, Bloomington, IN 47408 USA \\ \email{azadnema@indiana.edu} \and 
Nathaniel Rodriguez \at Center for Complex Networks and Systems Research, School of Informatics and Computing, Indiana University, Bloomington, IN 47408 USA,\\ \email{njrodrig@iu.edu} \and Alessandro Flammini \at Center for Complex Networks and Systems Research, School of Informatics and Computing, Indiana University, Bloomington, IN 47408 USA,\\ \email{aflammin@iu.edu} \and Yong-Yeol Ahn \at Center for Complex Networks and Systems Research,  School of Informatics and Computing, Indiana University, Bloomington, IN 47408 USA,\\ \email{yyahn@iu.edu}
}

\maketitle

\abstract{In this chapter, we apply the theoretical framework introduced in the previous
chapter to study how the modular structure of the social network affects the
spreading of complex contagion. In particular, we focus on the notion of
\emph{optimal modularity}, that predicts the occurrence of global cascades
when the network exhibits just the right amount of modularity. Here we
generalize the findings by assuming the presence of  multiple communities and
an uniform distribution of seeds across the network. Finally, we offer some
insights into the temporal evolution of cascades in the regime of the optimal
modularity.
\\
\textbf{Keywords} mean-field approximation, social contagions, community structure, linear threshold model, Watts threshold model, optimal modularity}

\section{Introduction}

The previous chapter reviewed the message-passing (MP) framework that can
accurately describe the dynamics of spreading processes, and in particular that
exhibited by the Watts threshold model~\cite{Granovetter1978threshold,
Watts2002simple, gleeson2008cascades}. In this chapter, we leverage the
framework to study how complex contagions are affected by the modular structure
of the underlying social network. In particular, we focus on the notion of
\emph{optimal modularity}, that predicts the occurrence of global cascades
when the network exhibits just the right amount of
modularity~\cite{nematzadeh2014optimal}. 

Modular organization, or community structure, is one of the most ubiquitous
properties of real-world networks~\cite{Girvan2002community,
newman2006modularity} and therefore it is crucial to understand how information
diffusion is affected by a modular structure. Addressing this problem is
particularly urgent when one considers  spreading phenomena characterized  by
complex contagion. Unlike the case of simple contagion, where modules simply
slow down the spreading, complex contagion may be either enhanced or hampered
by modular structure~\cite{Centola2010spread, Weng2013viral}. In contrast to
simple, complex contagion requires multiple exposures and those are favored
within densely connected communities. At the same time, complex contagion can
be strongly hampered at the boundaries of communities due to the lack of the
sufficient connectivity needed to provide the required multiple exposures from
the activated community to the yet-to-be-activated one.  The counter-intuitive
phenomenon of optimal modularity arises from the clash and compromise between
these opposite tendencies.

The basic setting for our study is as follows. We assume a network of
individuals where an individual can be in either an ``active'' or ``inactive''
state.  At each time step, an inactive node may become active if the node is
surrounded by \emph{enough} active nodes. The activation condition is captured
in a threshold function $C\left(m,k\right)$ that typically depends on the
degree $k$ of a node, and the number $m$ of its active neighbors. Here we
consider $C\left(m,k\right) = H(\frac{m}{k} - \theta)$, where $H(x)$ is a
Heaviside step function and $\theta$ is a threshold value. Throughout this
chapter we assume that $\theta$ is constant across the network. Our analysis
leverages the framework introduced in the previous chapter. We focus our
analysis only on the ensembles of random networks with arbitrary degree
distribution~\cite{newman2001random}, and ``message-passing'' (MP) and
``Tree-Like'' (TL) are used interchangeably throughout our chapter. 


\section{Analytical framework} 

\subsection{Mean-field and message-passing approaches for configuration model} 

As explained in the previous chapter, the steady-state fraction of active nodes
$\rho_\infty$ can be estimated using  Mean-Field (MF) or the Message-Passing
(MP) approaches. Assuming an underlying infinite networks with a given  degree
distribution $p_k$ but otherwise random, $\rho_\infty$ can be obtained by
solving the following self-consistent equations. Using the MF approach, 
\begin{equation}
\rho_\infty = \rho_0 + (1 - \rho_0)\sum_k p_k \sum_{m=0}^{k} {k \choose m}
(\rho_\infty)^m (1-\rho_\infty)^{k-m} C\left(\frac{m}{k}\right), 
\end{equation}
where $\rho_0$ is the initial fraction of seeds. This approach does not aim at
describing the evolution from one time step to the other, rather it states that
at stationarity, the density of active nodes is the sum of two contributions:
the fraction of seed nodes and expected number of nodes that have an
above-the-threshold fraction of active neighbors. This last contribution, in
turn, is expressed in terms of the degree distribution and of the density of
active nodes itself.

The MP (TL) approach assumes that the underlying network is well approximated
by a tree structure. To the extent to which such approximation is valid, where
each node is  affected only by its children. The density of active nodes
depends only from the level of the tree where the node is, which is described
by the following formula: 
\begin{equation}\label{eq:mp_tree} 
q_n = \rho_0 + (1 - \rho_0)\sum_k \frac{k}{\langle k \rangle} p_k
\sum_{m=0}^{k-1} {k-1 \choose m} (q_{n-1})^m (1-q_{n-1})^{k-1-m}
C\left(\frac{m}{k}\right), 
\end{equation} 
where $q_n$ is the density of active nodes at the $n$-th level of the tree
($q_0 = \rho_0$).  Note that \emph{excess degree distribution} is used in the
place of degree distribution because each node uses one of its link to connect
to its parent and only children nodes affect the status of the node.  The final
density can be calculated by focusing the root node:
\begin{equation}\label{eq:mp_root} 
\rho_\infty = \rho_0 + (1 - \rho_0)\sum_k p_k \sum_{m=0}^{k} {k \choose m}
(q_\infty)^m (1-q_\infty)^{k-m} C\left(\frac{m}{k}\right), 
\end{equation}
where $q_\infty = \lim_{n\rightarrow\infty} q_n$. See the previous chapter for
more details. 


\subsection{Generalization to modular networks} 

The MP framework can be readily generalized to modular networks by introducing
density-of-active variables for each community ~\cite{gleeson2008cascades}.
Consider a network with $d$ communities, where the connection probabilities
between communities are stored in a $d \times d$ mixing matrix $\mathbf{e}$.
Here $e_{ij}$ is the probability that a random edge connects community $i$ and
$j$. Consider a node in community $i$ and at the $n+1$ level of the spreading
tree. The probability to pick one of its active children ($n$-th level) can be
written as:
\begin{equation}\label{MP1}
\overline q_n^{(i)} = \frac{\sum_j e_{ij} q_n^{(j)}}{\sum_j e_{ij}}. 
\end{equation}

\noindent Eq.~\ref{eq:mp_tree} can be extended to describe the relation between
the densities in different communitites~\cite{gleeson2008cascades}. 
\begin{align}\label{MP2}
q_{n+1}^{(i)} = \rho_0^{(i)} + (1-\rho_0^{(i)}) \sum_k \frac{k}{z^{(i)}}
\sum_{m=0}^{k-1} {k-1 \choose m} (\overline q_{n-1}^{(i)})^m (1- \overline
q_{n-1}^{(i)})^{k-1-m} C^{(i)}\left(\frac{m}{k}\right),
\end{align}
\noindent where $z^{(i)} = \sum_k k p_k^{(i)}$ is the mean degree of community
$i$. This set of  equations can be solved by iteration analogously to
Eq.~\ref{eq:mp_root}. The  density of active nodes at stationarity is: 
\begin{equation}
\rho_\infty = \sum_i \frac{N^{(i)}}{N} \rho_\infty^{(i)}. 
\end{equation}
See~\cite{gleeson2008cascades} for more details.



\section{Networks with two communities} 

Having set up the necessary tool, we now turn into investigating on how the
strength of the modular structure can affect the spreading of complex
contagion.  The simplest setting one may consider is a network with two equally
sized communities.  Given a fixed and predefined number $L$ of links in the
network, we first randomly connect $\mu L$ couple of nodes, where each member
of the couple sits in a different community. The remaining links are then used
to randomly connect couple of nodes in the same
community~\cite{Girvan2002community}. If $\mu=0$, no edge is placed between the
two communities (the network has two components and is therefore maximally
modular); if $\mu=0.5$, the network is an Erd\H{o}s-R\'enyi random graph in the
infinite size limit. A fraction $\rho_0$ of active nodes are set in one of the
two communities, which we call ``seed community''. 

\begin{figure}[b!] 

\includegraphics[width=\columnwidth, clip=true]{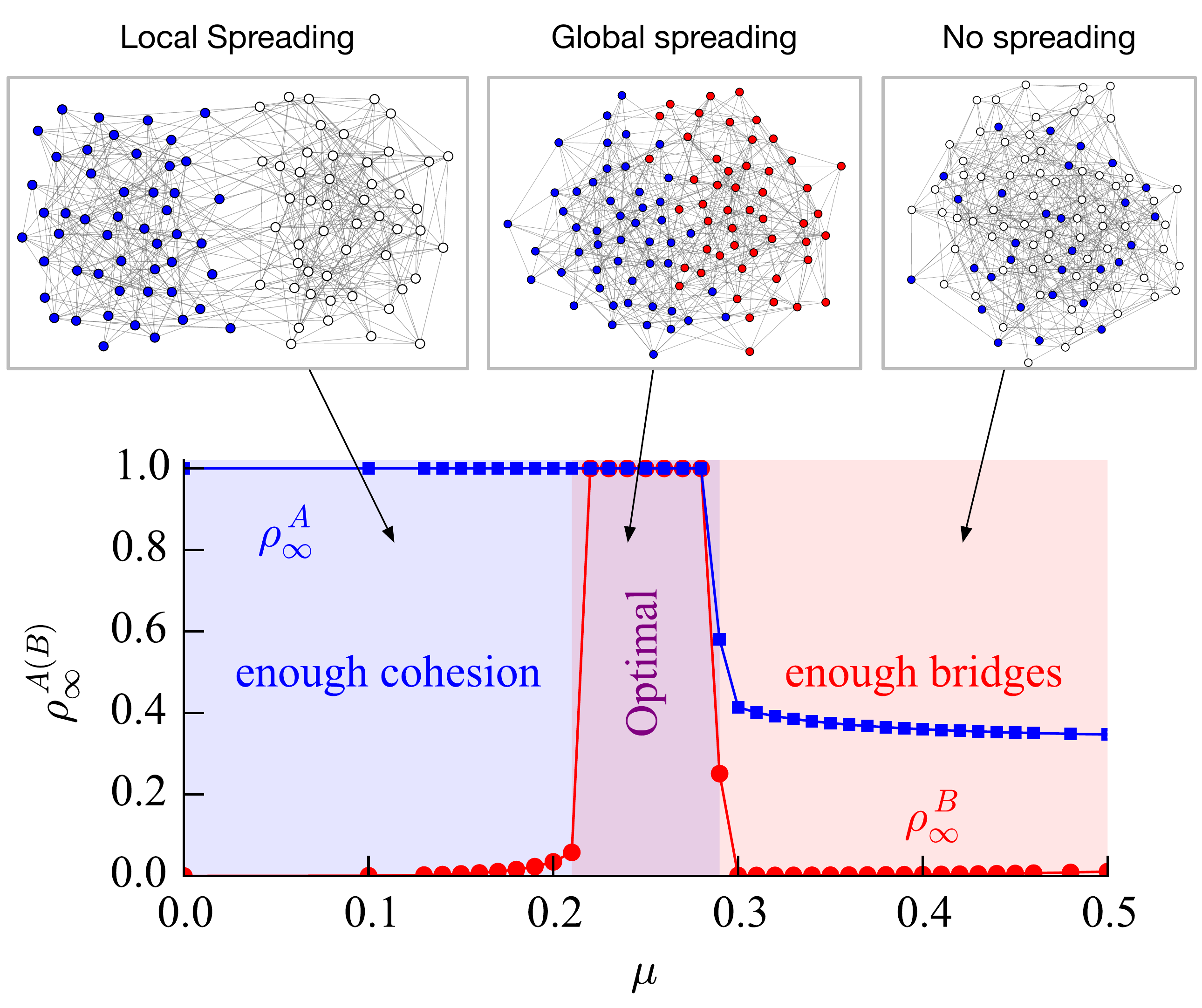}

\caption{The trade-off between intra- and inter-community spreading.  Stronger
communities (small $\mu$) facilitate spreading within the originating community
while weak communities (large $\mu$) provide bridges that allow spreading
between communities. Blue and Red imply activation, while white implies
inactivity.  There is a range of $\mu$ values that allow both (optimal).  The
blue squares represents $\rho_\infty^A$, the final density of active nodes in
the community $A$, and the red circles represents $\rho_\infty^B$.  The
parameters for the simulation are: $\rho_0 = 0.17$, $\theta = 0.4$, $N =
131056$, and $\langle k \rangle = 20$. }\label{fig:optimum_mu}

\end{figure} 

As shown in~\cite{nematzadeh2014optimal} and  illustrated in
Fig.~\ref{fig:optimum_mu}, the density of active nodes at stationarity
$\rho_\infty$ depends non-trivially on the degree of inter-community
connectivity, showing a maximum at intermediate values of $\mu$.

Small values of $\mu$ allow  initial spreading in the seed community, but it is
essential to have enough mixing (bridges) between communities to have a cascade
that significantly interests the global community. At the same time, when too
many links across community are present, since these occur at the expense of
the intra-community links, there is insufficient connectivity in the seed
community to trigger the initial diffusion of the activation. We name
\emph{optimal modularity} the range of $\mu$ values for which the two
mechanisms above find their trade-off to maximize the size of the cascade. 


\section{Optimal modularity in networks with many communities} 

A network with just two equally sized communities with all the seed users
concentrated in one of those is an obvious starting point for this study, but,
in general, a non-realistic assumption.  We generalize our finding by first
considering  multiple communities of the same size and with the same degree of
intra and inter-connectivity. We then consider a more general process to
generate the network and its modular structure. We consider the family of
graphs known as LFR (Lancichinetti-Fortunato-Radicchi) benchmark
graphs~\cite{lfr2008bench}, which allow us to independently modulate both the
size and the degree distribution of the individual communities. We finally
remove the constraint of having all seed nodes in a single community.

\subsection{Spreading from a seed community}

\begin{figure} 
\centering
\includegraphics[width=\columnwidth, clip=true, trim=0 10 0 8]{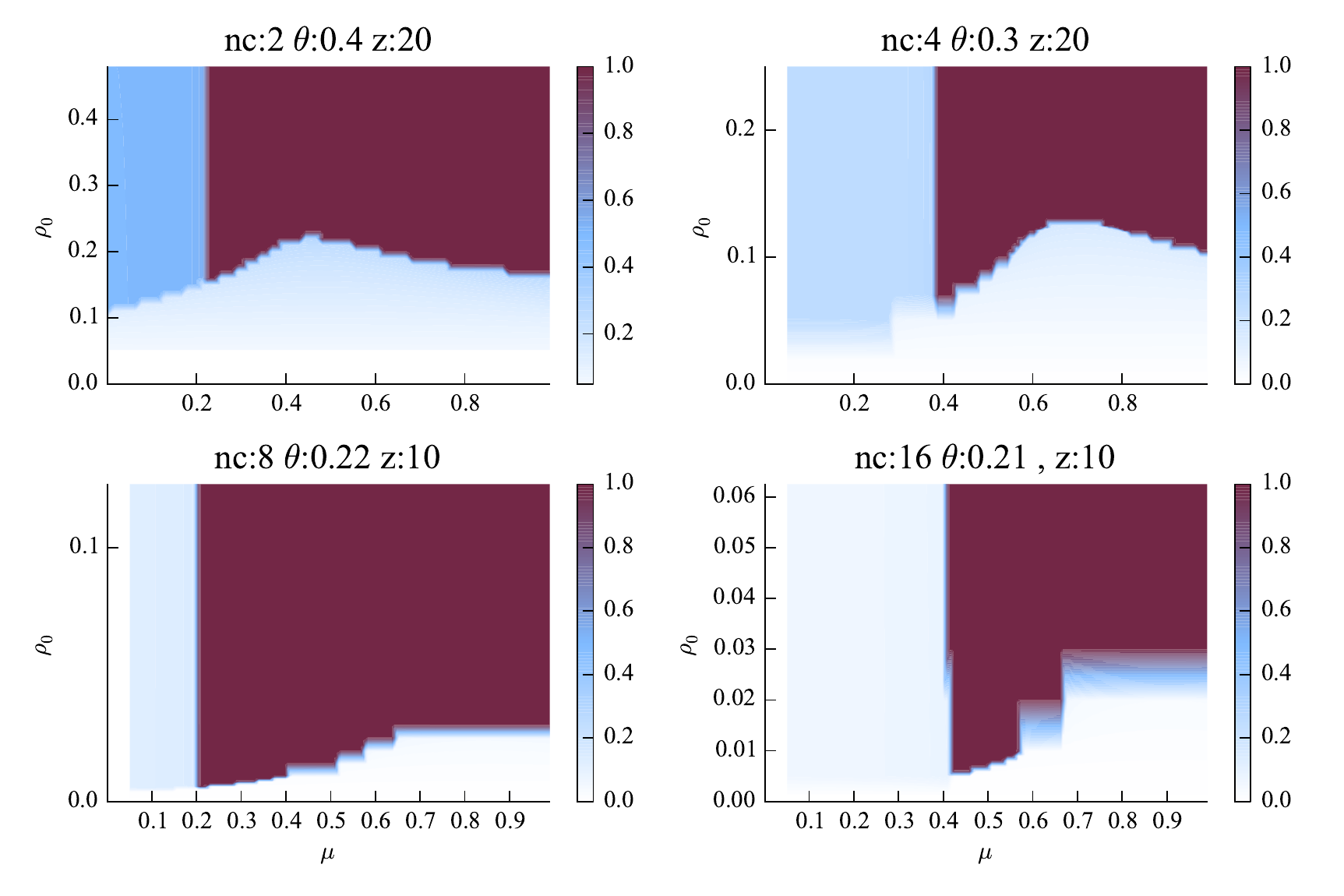}

\caption{Optimal modularity arises even when there are many communities. The
phase diagrams are calculated using the MP framework with different number of
communities. `nc' refers the number of communities. $\theta$ and $z$ values are
varied to demonstrate the existence of optimal modularity clearly.}
\label{fig:fig_commu}

\end{figure} 

\begin{figure} 
\centering
\includegraphics[width=0.7\columnwidth, clip=true, trim=0 10 0 8]{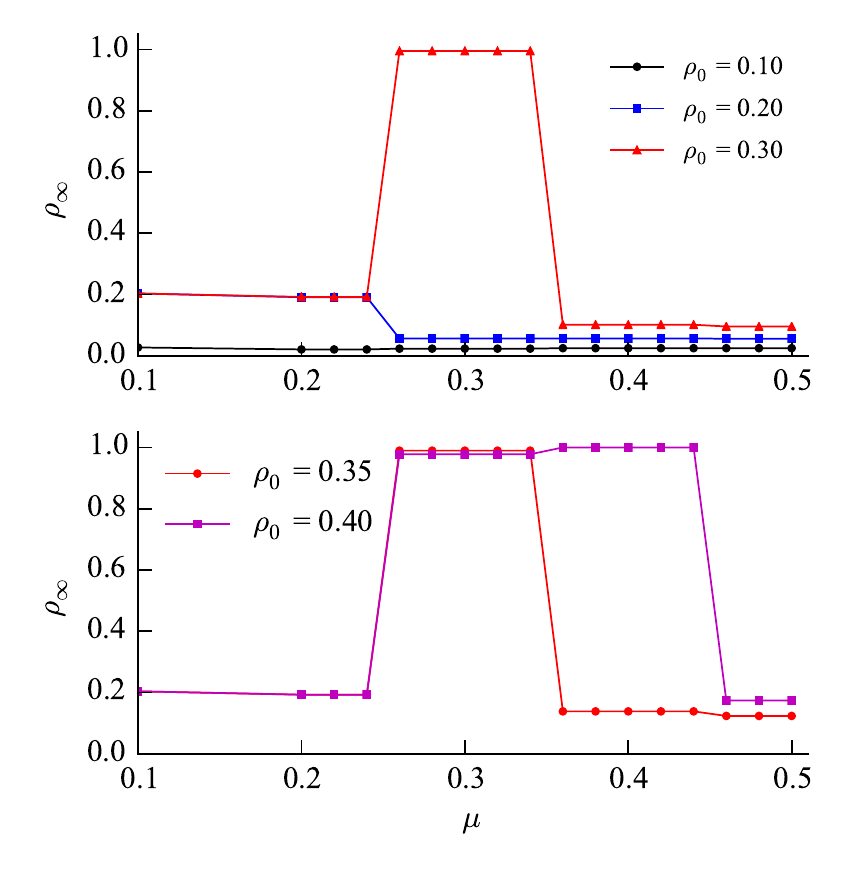}

\caption{The behavior of threshold model in the presence of community
structures generated by LFR benchmark, with $N=25000$, $z=10$, $t_1=2.5$
(degree exponent), $t_2=1.5$ (community size exponent), $k_{max}=30$ and
$\theta=0.3$.  LFR benchmark generates more \emph{realistic} networks with
community structures. The degree distribution may have a power-law distribution
(with exponent $t_1$ and degree cutoff $k_{max}$). The size of the communities
may also follow a power-law distribution (with exponent $t_2$).
\label{fig:fig_lfr}}

\end{figure}

Figure~\ref{fig:fig_commu} and~\ref{fig:fig_lfr} show that the  qualitative
behavior of $\rho_\infty$ as a function of $\mu$ and $\rho_0$, when there are
two or more communities and the activation is iniated from a single seed
community.  In general, a larger number of communities requires a smaller
adoption threshold to allow the cascade to spread over the all network;
increasing the number of communities makes the signal outgoing from the seed
community less focused. Such signal, therefore, spreads less easily from
community to community. Nevertheless, the same trade-off, and thus optimal
modularity, exists between local spreading due to clustering and
inter-community spreading due to bridges. 

\subsection{Spreading from randomly distributed seeds}

Next we consider the more general scenario in which the initial signal is
distributed across the whole network.  The MP approach as described by
Eqs.~\ref{MP1} and \ref{MP2} is sufficiently general to handle the scenario at
hand. In particular we will have:
\begin{eqnarray}
\bar{q}_n^{(i)}=(1-\mu)q^{(i)}_n+\frac{\mu }{(d-1)} \sum_{j\neq
i}^{d}q^{(j)}_n.
\end{eqnarray}
\noindent Here $d$ is the number of communities and, as before, $\mu$
represents the total fraction of inter-community bridges in the network. Also,
in Eq~\ref{MP2}, $\rho_0^{(i)} \neq 0$  for all $i$'s rather than just one.
The equations can still be solved iterativly.

\begin{figure} 
\centering
\includegraphics[width=\columnwidth, clip=true, trim=0 0 0 0]{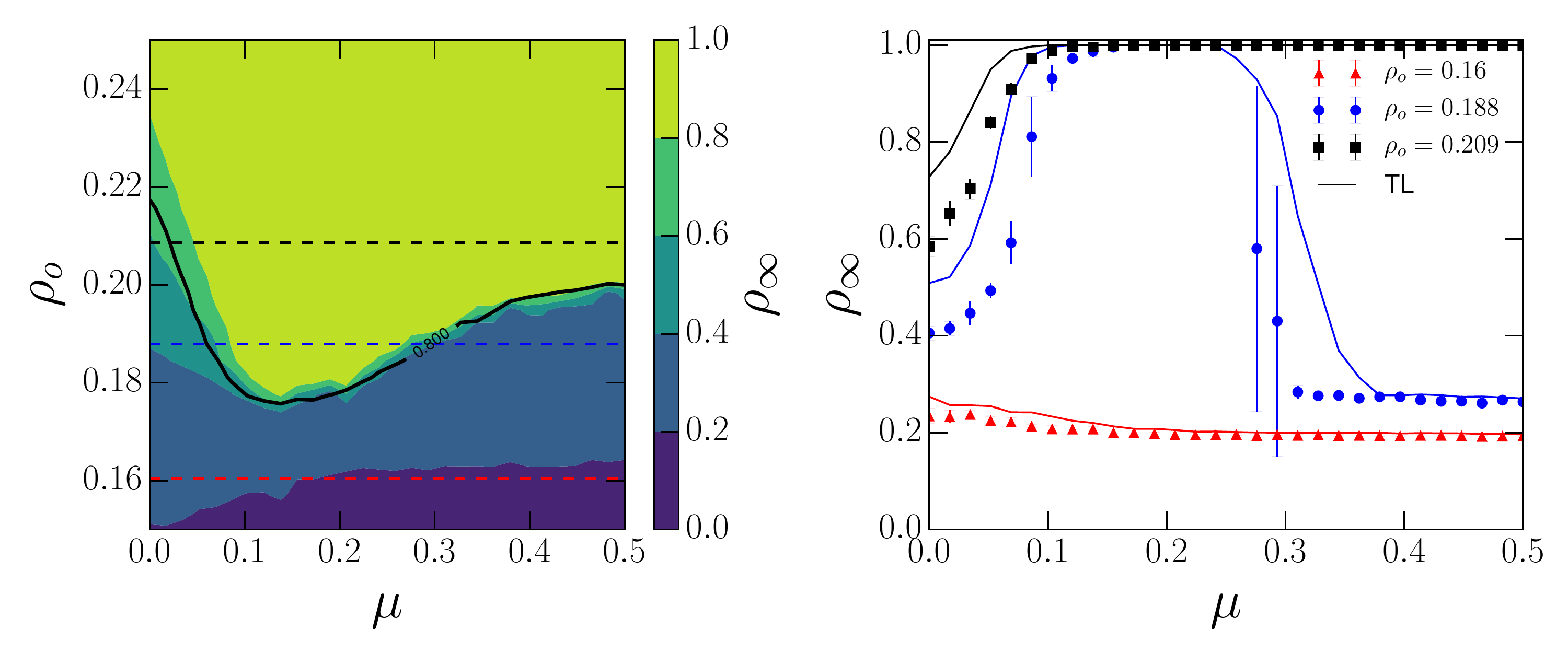}

\caption{The phase diagram of threshold model with uniformly distributed random
seeds. Three example slices are taken from the contour plot (horizontal lines)
and displayed in the right figure. $N=25,600$ with $C=160$ communities with
$160$ nodes each. The solid black line on the contour shows the MP (TL)
results.} \label{fig:multicom_mu_v_seed}

\end{figure}

Figure~\ref{fig:multicom_mu_v_seed} shows the results derived via the MP (TL)
approach for a network with $25,600$ nodes, $160$ evenly sized communities, and
seeds randomly distributed across the network. An optimal region emerges as in
the previous multi-community cases. 

We would like to note that the optimal region vanishes if each community has
exactly same number of seeds; there is no dependence of $\rho_\infty$ on $\mu$.
The emergence of an optimal region critically depends on the existence of
variability across communities.  Individual communities show a sharp transition
between inactivity and activity as the seed fraction $\rho_o$ increases.  As
all nodes activate essentially simultaneously, the entire system can be
regarded as a random network of super-nodes, each representing a single
community. The qualitative behavior is therefore the same as that of a random
network with no communities.  If there is sufficient variability across the
communities, in terms of the number of seed nodes they contain then some
communities will activate before others and the community structure will have a
measurable effect, as shown in Fig~\ref{fig:multicom_mu_v_seed}.


The effect of variability in the nodes' threshold was actually the focus of the
original study of the linear threshold model by
Granovetter~\cite{Granovetter1978threshold}. He found that changes in the
variance of the threshold distribution leads to qualitatively different
spreading behavior even when the mean is kept constant. When variance is low,
nodes have approximately the same threshold and, all other factors being the
same, they get activated more or less simultaneously. Higher variance brings
the existence of a continuum spectrum from low to high threshold nodes. Low
threshold are typically activated first and can help the activation of nodes
with slightly higher threshold. In turn, these can contribute to activate even
higher threshold nodes, and possibly generate a large size cascade. But if
threshold variance is too high, the gap between low threshold and high
threshold nodes is too large and the activation of the former is not sufficient
to fill the gap in threshold.

Given the tendency of nodes in a community to activate simultaneously, it is
possible to regard them as coarse-grained super-nodes whose threshold is
effectively determined by the number of seeds they contain. This formulation
provides similar insights as the Granovetter's
study~\cite{Granovetter1978threshold}.  We investigated this idea by
distributing seeds across communities according to a Beta distribution. We fix
the $\alpha$ and $\beta$ parameters for the Beta distribution in such a way to
maintain the expected value constant at $\langle \rho_o \rangle=0.19$ while the
standard deviation ($\sigma$) is varied. 

\begin{figure}
\centering
\includegraphics[width=\columnwidth, clip=true, trim=0 0 0 0]{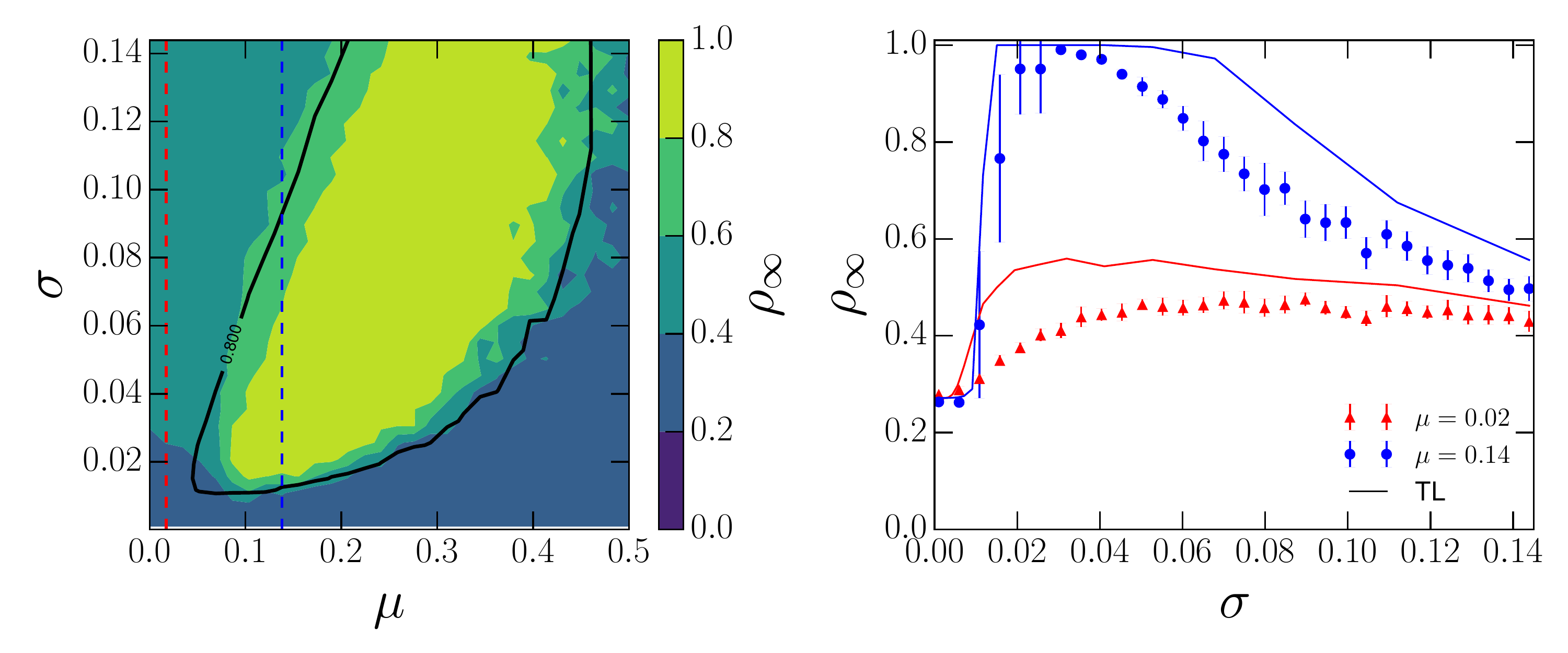}
\caption{The phase diagram of threshold model with beta distributed random seeds. Two example slices are taken from the contour plot (vertical lines) and displayed in the right figure. The mean seed $\langle \rho_o \rangle=0.19$. Numerical simulations were done with $N=25600$ with $C=160$ communities with $160$ nodes each. The solid black line on the contour shows the MP results.} 
\label{fig:multicom_mu_std}
\end{figure}

Our experiments, as shown in Figure~\ref{fig:multicom_mu_std} produce results qualitatively similar to those for Granovetter's model~\cite{Granovetter1978threshold}. 

Specifically, Fig.~\ref{fig:multicom_mu_std} shows that two optimal behaviors
emerge, one with respect to $\sigma$ and one with respect to $\mu$. The peak
along $\sigma$ arises for exactly the reasons exposed above. A large cascade
can be triggered for intermediate values of $\sigma$, when there is a continuum
spectrum of effective activation thresholds across communities.  Due to a
cascading effect, increasing activity within the network makes it more likely
to activate communities with fewer and fewer seeds. When $\sigma$ is low
communities have roughly the same number of seeds and none have enough seeds to
fully activate unless the mean number of seeds is increased. At high $\sigma$,
communities with many seeds activate, but they don't generate enough cumulative
activity to activate the low seed communities.

The optimal region with respect to $\mu$ arises for the same trade-off to those
studied above. Low $\mu$ implies strong connectivity inside single communities,
but insufficient bridges to spread the activation signal externally. For high
$\mu$ there are bridges, but not sufficient internal connectivity in order to
trigger the initial activation of a sufficiently large number of communities.
An optimal balance is achieved at intermediate values of $\mu$.




\section{Temporal aspects of optimal modularity} 

 \emph{How fast} a contagion can spread is often as important as \emph{how far}
it can spread. Imagine, for example, the sudden availability of a prophylactic
measure in the wake of a pandemic. The issue would then be not just whether
this measure can spread broadly, but also whether it can spread sufficiently
fast to effectively oppose the pandemic.  Here we limit our study to the basic
setting  consisting of two communities with varying degree of modularity
($\mu$) and only one seed community. We measure the total diffusion time:  the
number of time steps needed for the system to reach a steady state. We run a
1,000 simulations (each with an independent network realization) and measure
the mean $\rho_{\infty}$ and total diffusion time. We also assume a uniform
threshold ($\theta=0.4$). 

\begin{figure*} 
\centering
\includegraphics[width=\columnwidth, clip=true]{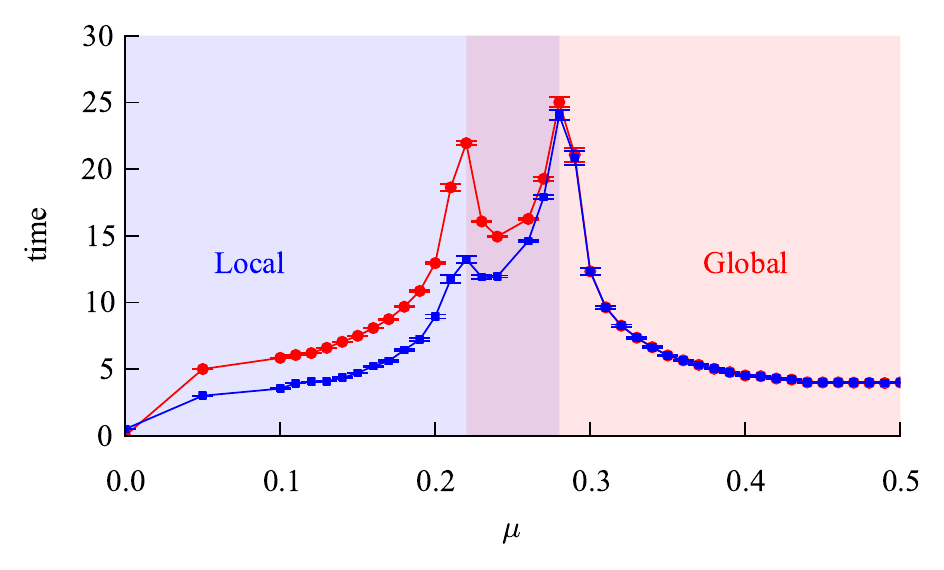}

\caption{Total diffusion time and optimal modularity. The blue symbols and line represent the total diffusion time in the community $A$ (seed community), and the red symbols and line represent the total diffusion time in the community $B$ (the other community). The optimal modularity range that allows global cascades is represented with a purple shade. The total diffusion time curve peaks at the two transition points, demonstrating that there exists a narrower range of $\mu$ values where the global cascades happen, faster. The parameters for the simulation are: $\rho_0 = 0.17$, $\theta = 0.4$, $N = 8,192$, and $z = 20$. }
\label{fig:optimum_mu_time}

\end{figure*} 

Figure~\ref{fig:optimum_mu_time} demonstrates that while
$\rho_\infty$ remains constant at its maximum value,  the total diffusion time
greatly varies.  Close to either border of the optimal range, contagion
significantly slows down, while the global spreading can happen fastest near
the middle of the optimal modularity regime. When there are just enough bridges
(the left border), the spreading from the seed community to the other community
is slower than the case where there are more than just enough bridges to spare
(center).  Similarly, when there are just enough local cohesion (the right
border), the local spreading produces just enough newly activated nodes to
achieve global cascade, slowing down the spreading process.


\section{Discussion} 

In this chapter, we have generalized the optimal modularity phenomena and
studied its temporal aspect.  We showed that many simplifying assumption made
in the original study can be relaxed without disrupting the qualitative
scenario that predicts a maximum in the fraction of active individuals for
intermediate values of inter-community connectivity. In particular we
considered the case of a large number of communities, with heterogeneous size,
and non uniform degree of initial activation.

Our experiment showed that our model behaves qualitatively same as one in which
communities can be considered as super-nodes and are characterized by different
threshold. This, in turn, may open the possibility to study very large system
if one could devise a strategy to compute the effective parameters of a coarse
grained model where communities are represented by single nodes.  The interest
in developing such "renormalization" techniques is not only theoretical:
threshold models have found several application to real world problems,
including the multi-scale modeling of brain networks~\cite{sporns2004brain} and
of their activation dynamics in the brain~\cite{hilge2011hmn}. 


\bibliographystyle{unsrt}
\bibliography{refs}

\end{document}